# Phase retrieval of an electron vortex beam using diffraction holography

Federico Venturi [1, 2], Marco Campanini [3,**], Gian Carlo Gazzadi [2], Roberto Balboni [4], Stefano Frabboni [1, 2], Robert W. Boyd [5, 6], Rafal E. Dunin-Borkowski [7], Ebrahim Karimi [5, 8], Vincenzo Grillo [2, 3 *]

[1] Dipartimento FIM, Università di Modena e Reggio Emilia, Via G. Campi 213/a, I-41125 Modena, Italy
[2] CNR-Istituto Nanoscienze, Centro S3, Via G. Campi 213/a, I-41125 Modena, Italy
[3] CNR-IMEM Parco Area delle Scienze 37/A, I-43124 Parma, Italy
[4] CNR-IMM Bologna, Via P. Gobetti 101, 40129 Bologna, Italy
[5] Department of Physics, University of Ottawa, 25 Templeton, Ottawa, Ontario, K1N 6N5 Canada
[6] Institute of Optics, University of Rochester, Rochester, New York 14627, USA
[7] Ernst Ruska-Centre for Microscopy and Spectroscopy with Electrons and Peter Grünberg Institute, Forschungszentrum Jülich, 52425 Jülich, Germany
[8] Department of Physics, Institute for Advanced Studies in Basic Sciences, 45137-66731 Zanjan, Iran

*corresponding author: vincenzo.grillo@cnr.it

** present address: EMPA, Überlandstrasse 129, 8600 Dübendorf (Zürich)

In both light optics and electron optics, the amplitude of a wave scattered by an object is an observable that is usually recorded in the form of an intensity distribution in a real space image or a diffraction image. In contrast, retrieval of the phase of a scattered wave is a well-known challenge, which is usually approached by interferometric or numerical methods. In electron microscopy, as a result of constraints in the lens setup, it is particularly difficult to retrieve the phase of a diffraction image. Here, we use a "defocused beam" generated by a nanofabricated hologram to form a reference wave that can be interfered with a diffracted beam. This setup provides an extended interference region with the sample wavefunction in the Fraunhofer plane. As a case study, we retrieve the phase of an electron vortex beam. Beyond this specific example, the approach can be used to retrieve the wavefronts of diffracted beams from a wide range of samples.

In light optics and electron microscopy, the intensity distribution in a real space image or a diffraction image of a sample is usually recorded, while phase information is lost. For this reason, significant effort has been aimed at developing methods that can be used to retrieve phase distributions. Retrieval of the phase from both real space and diffraction images is particularly important in electron microscopy, as it is sensitive to built-in electric and magnetic fields in materials, crystallographic structure and local strain, as well as providing a route to enhancing contrast from weakly scattering objects [1-4]. A recent addition to the list of objects, for which characterization of the phase is important comes from the formation of electron vortex beams (EVBs), which possess orbital angular momentum (OAM) $L_z = \ell\hbar$ with topological charge $\ell$, where $\hbar$ is the reduced Plank constant [5,6]. EVBs are usually generated by diffraction from nanofabricated computer-generated holograms (CGHs) and are currently receiving considerable attention for both fundamental research and applications [7-9]. A key difficulty in phase retrieval of EVBs results from the fact that they possess phase singularities in the electron beam transverse plane [10].

All of the methods that are used to retrieve the phase of a wavefront can be related to the original idea of Gabor of electron holography [11]. He proposed that the relative phase of a sufficiently coherent wave system can be retrieved by interference with a known reference wave. Depending on the angle between the reference wave and the normal to the hologram plane during the recording step, holographic methods can be divided into either in-line (parallel) or off-axis (tilted) modes. The in-line approach is an elegant, yet computationally demanding, approach to phase reconstruction. It is based on the recording of a number of images, either at different defocus values or in real and diffraction space. The phase retrieval processes is then either iterative, as in the case of the Gerchberg–Saxton algorithm, or deterministic, as in the case of the transport of intensity equation (TIE) [12]. An in-focus version of the real space in-line holography method, which is potentially very powerful but is still under development, is based on the use of electron phase plates [13]. A more widely used and direct method of phase retrieval in an electron microscope is based on the interference of a wavefunction of interest with a tilted reference wave. This approach typically involves using an electron biprism [14] to split a single wavefront into two different nearly-plane wavefronts, only one of which passes through the sample, in order to produce an electron interference pattern [15,16] or an off-axis electron hologram [17-20]. The disadvantages of this approach include the limited size of the interference region that can be formed and the need for a perfectly unperturbed reference wave. Moreover, the interference pattern is not recorded directly in the biprism plane, but some distance from it, in order to allow superposition. A comparison between in-line and off-axis electron holography has recently been reported [21].

Phase retrieval methods are equally important for diffraction images, which can be recorded from a large sample area and have many applications, especially for X-rays, where the wavelength and lensing make imaging more difficult (see, e.g., [22-24]). Phase retrieval methods that are based on iterative approaches have also been applied to nanobeam electron diffraction [25] in both high-resolution [26] and low-resolution [27,28] scanning transmission electron microscopy (STEM). In electron crystallography, high-resolution real space images and diffraction images have been combined to solve the crystallographic phase problem [29].

For an EVB, phase retrieval from a diffraction image of a CGH is required. Previously, accurate real space characterization of a CGH was simply used to infer the phase of an EVB [30-32]. Interference in the out-of-focus Fresnel regime has also been carried out [33]. However, a direct measurement of the phase of an EVB would be preferable, in particular when such wavefunction is produced by elastic electron-sample interaction. Unfortunately, the standard experimental configuration of the biprism-based approach to electron holography cannot be transferred directly to diffraction, as the Fraunhofer diffraction pattern of the reference wave would simply be an image of the source, i.e., ideally a point. In addition, it has been demonstrated that TIE, in its simplest implementation, cannot be used to provide a map of vortex singularities, but instead requires a complicated generalization [34] or astigmatic illumination, as reported for X-ray optics [35].

The aim of the present paper is to demonstrate that, by using CGHs and holographic reconstruction methods, it is possible to combine the advantages of large area illumination and in-focus interference to achieve phase retrieval in the diffraction plane, thereby enabling phase retrieval of an EVB.

In order to obtain an appropriate interference pattern in the diffraction plane, it is necessary to structure the reference beam in a controlled manner. A simple case such as a Bessel beam [31,36], whose CGH is characterized by a linear phase gradient in the radial direction, only produces rings. In contrast, a defocused beam (DB) is associated with a quadratic variation in phase and produces an extended diffraction pattern that can be superimposed onto the diffraction pattern of an object. This concept is similar to diffraction holography in X-ray experiments [37,38]. However, in our case the technology that is used to produce nanofabricated holographic phase plates for the generation of EVBs can also be used to structure the reference wave.

In the present study, two closely-spaced CGHs are fabricated using focused ion beam (FIB) milling on a single SiN membrane following the same procedure that is normally used to produce phase holograms [30-32]. One of the holograms is the DB-CGH, which has a parabolic phase-modulation so that it produces a reference DB. The other hologram is the EVB-CGH, which is used to generate an EVB with an OAM $L_z$ of 10 $\hbar$. The phase that is imprinted on the DB-CGH takes the form

$$\varphi(\rho,\theta) = \pi \left(1+\frac{1}{2}\text{sign}(\sin(k_{carr}\rho\sin\theta + a\rho^2))\right), \qquad (1)$$

where $\rho, \theta$ are polar coordinates, $k_{carr}$ is the carrier frequency and $a$ is a real value that is related to the curvature of the reference DB. Such a parabolic phase hologram also produces a parabolic wavefront in the Fraunhofer plane. This can be regarded as an extension of the Fourier transform of a Gaussian function [1].

As the desired shape of the vortex beam is a Laguerre Gauss [39,40] beam, we used a complicated encoding of phase and amplitude information [39,40] in the CGH. For the sake of discussion, we assume here that the phase imprinted in the EVB-CGH takes the form

$$\varphi'(\rho,\theta) = \mod((k_{carr}\rho\sin\theta + \ell\theta), 2\pi) \quad , \tag{2}$$

where the function *mod(m, n)* is the remainder of m on division by n and $k_{carr}$ is chosen to be the same for both holograms so that the centers of the first diffraction orders coincide. The value of $\ell$ was set to 10.

Images of the two CGHs recorded in a scanning electron microscope (SEM) are shown in Fig. 1a, while a sketch of the electron-optical setup in the TEM is shown in Fig. 1b. For comparison, Fig. 1c shows the setup for a plane reference wave, which produces no significant overlap with the object beam in the focal plane.

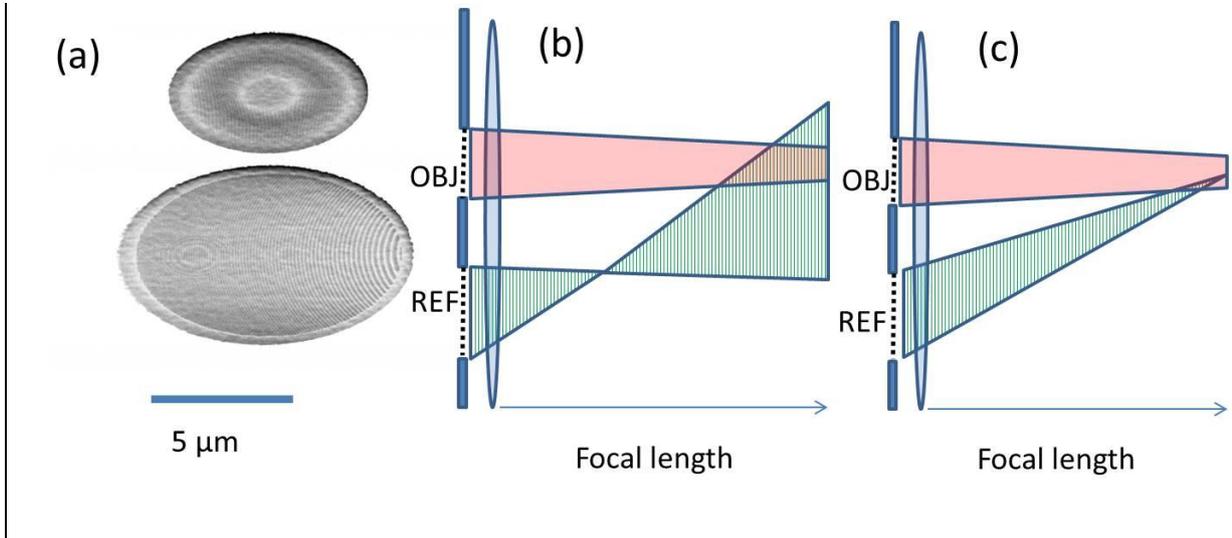

FIG. 1: (a) SEM images of the EVB-CGH (top) and the DB-CGH (bottom) recorded at an angle of 52°. (b, c) Simplified ray diagrams of interference in the focal plane of an object wave (OBJ) diffracted by the EVB-CGH with a reference wave (REF) that is either a DB diffracted by the DB-CGH (b) or a plane wave (c). The objective lens aperture that is used to obtain the diffraction image of each hologram has been omitted for clarity. The plane reference wave transforms into a point in the back focal plane and therefore cannot be used for interferometry.

Experiments were carried out at 200 kV in a JEOL 2200 TEM equipped with a Schottky field emitter. The microscope was operated in "LOW MAG" mode, with the "objective minilens" used as the main lens [30-32]. In this electron-optical configuration, we were able to observe diffraction images from the holograms *separately* by positioning the objective lens aperture (OLA) almost exactly in a plane conjugate to the specimen. Since high lateral coherence is required, nominally covering the full area of the two holograms (on the order of 20 µm), we optimized the illumination system to achieve highly parallel illumination, with the C3 condenser lens almost at its maximum excitation. When the OLA was removed, the diffraction images from the two holograms could be superimposed. The centers of both diffraction images were coincident, while the first diffraction orders were concentric. In detail, the centers of the two CGHs were separated by a distance D = 9.9 μm, while the diameters of the DB-CGH and EVB-CGH were 9.8 and 6.5 μm, respectively.

Diffraction images generated by the DB-CGH and EVB-CGH are shown in Figs 2c and 2g, respectively, while corresponding simulations are shown in Figs 2d and 2h. The DB exhibits circular diffraction, while the EVB shows circular symmetry and a dark region in the central area. As a result of hologram imperfections, the intensity is reduced at higher frequencies, i.e., on the lower side of Fig. 2c, which can be compared with the simulation in Fig. 2d. The curvature and size of the reference "defocused beam" are chosen so that the intensities of the first order diffracted beams are comparable in the superposition region.

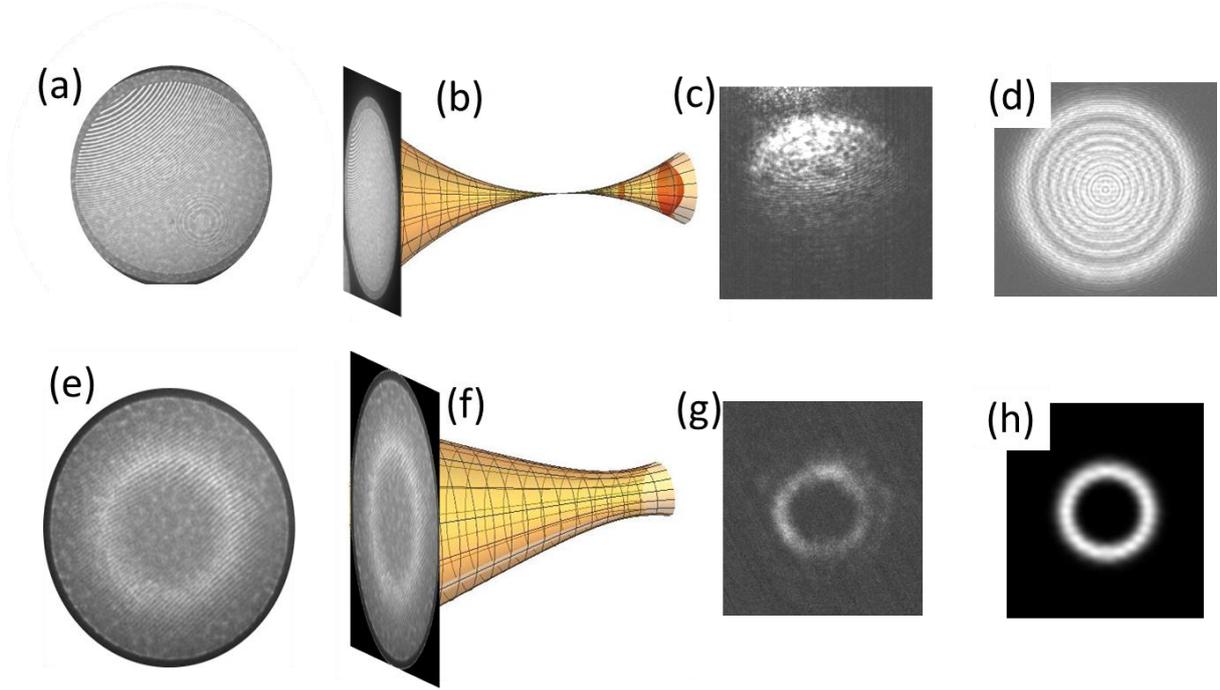

FIG. 2: TEM images of the DB-CGH (a) and of the EVB CGH (e) along with an illustration of the propagation (b,f). The fig c and g show the experimental diffraction patterns in the exact Fraunhofer plane. Clearly the diffraction from the DB is extended allowing a large superposition with the EVB. Fig. c and f show the corresponding simulations.

For the first order diffracted beam alone, in the Fraunhofer diffraction plane with in-plane coordinates $\bar{k}_\perp$ the electron wavefunction of the DB reference beam can be written in the form

$$\psi_{DB}(\bar{k}_\perp) = A \int exp(i\, a(\bar{\rho} + \bar{D})^2) \exp(i\overline{k_\perp} \cdot \bar{\rho})\, d\bar{\rho} \, , \qquad (3)$$

where the origin of the x,y coordinates is taken to be at the center of the EVB-CGH, while the center of the reference DB-CGH is located at $\bar{D}$. The wave that results from superposition of the two reconstructed waves takes the form

$$\psi_{DB}(\bar{k}_\perp) + \psi_{EVB}(\bar{k}_\perp) = A\, exp\left(i\frac{\pi^2 \bar{k}_\perp^2}{a}\right) \exp(i\overline{k_\perp} \cdot \overline{D}) + B exp(i\phi)\,, \tag{4}$$

where $\psi_{EVB}$ is the EVB wavefunction (or in general any wavefunction whose phase $\phi(\bar{k}_\perp)$ needs to be retrieved) and A and B are the amplitude of the two diffracted waves. The full interference pattern can be written

$$I(\bar{k}_\perp) = A^2 + B^2 + 2AB cos\left(\frac{\pi^2 \bar{k}_\perp^2}{a} + \phi(\bar{k}_\perp) + \bar{k}_\perp \cdot \overline{D}\right). \tag{5}$$

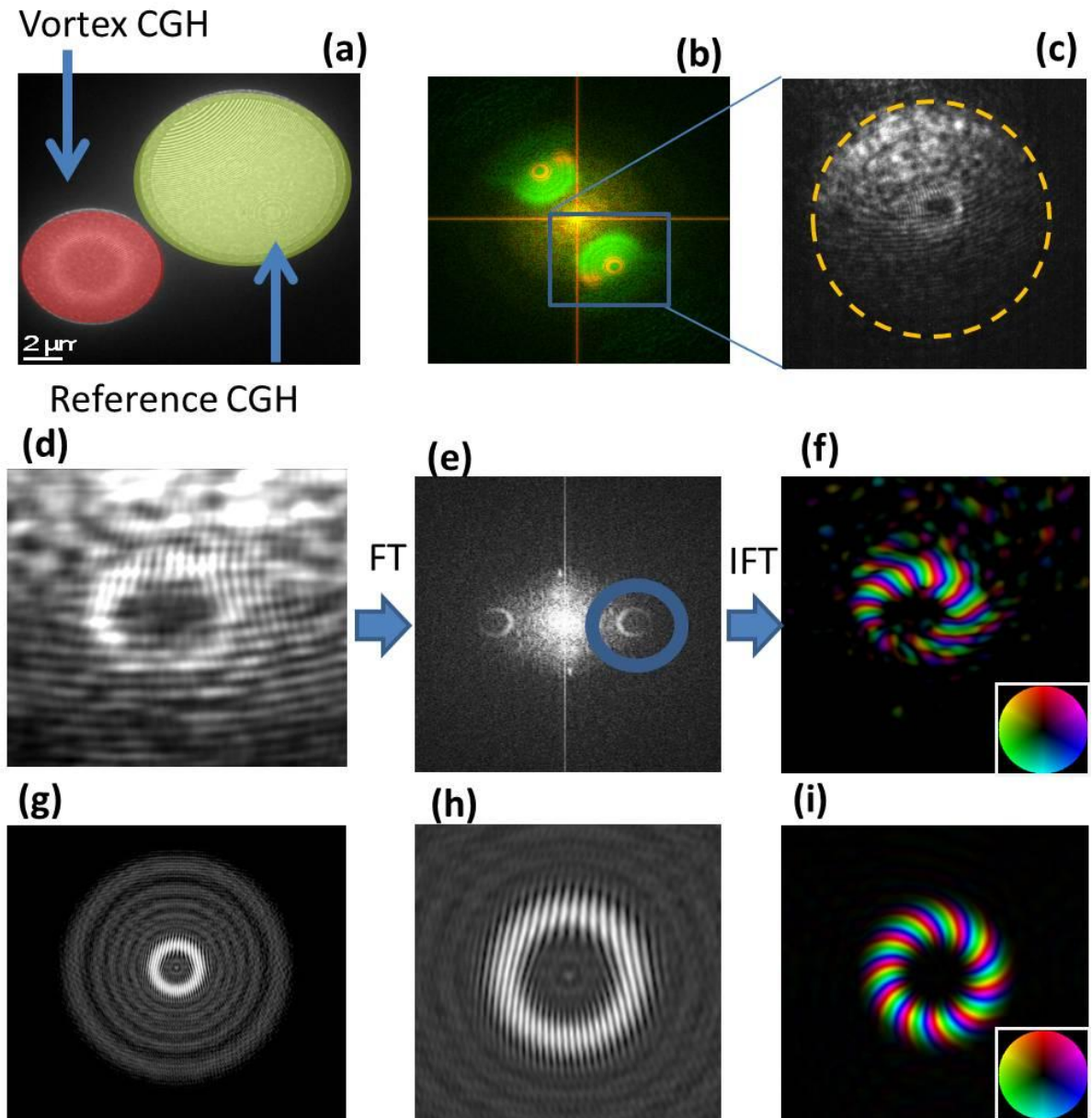

FIG. 3 Two fabricated CGHs in (a) give rise to a superposed diffraction pattern in (b), in which each contribution is shown using a different color. The first diffraction order, where superposition takes place, is shown in (c, d). The phase reconstruction algorithm considers region (d), where fringes arise, and involves the use of a digital Fourier transform (e), a translation of one of the sidebands to the origin of the reciprocal plane, and an inverse

Fourier transform (f). Phase and intensity information is shown in (f) using the hue and the saturation of the colors, respectively. Simulations of the superposed beams (g, h) and the reconstruction (i) are also shown.

The successive experimental steps are shown in Fig. 3. The interference pattern is visible in Figs. 3c and 3d. In the present case, we expect

$$\phi(\bar{k}_\perp) = \ell \operatorname{atan}\left(\frac{k_y}{k_x}\right). \tag{6}$$

Just as for an isolated DB, nearly horizontal fringes are visible due to the zero order tails superimposing onto the first order diffracted beam. However, these fringes are nearly orthogonal to the interference fringes that carry the EVB phase information and can be distinguished easily in the reconstruction algorithm.

Apart from the parabolic phase term, Eq. (5) resembles the interference effect in biprism-based off-axis electron holography. We can therefore apply the well-known algorithms for phase reconstruction that are used in those experiments. The steps are shown in the lower part of Fig. 3. The digitally calculated discrete Fourier transform (FT) of the intensity can be written in the form

$$FT(I) = (A^2 + B^2)\delta(\bar{q}) + AB\ FT\left[\exp\left(i\frac{\pi^2 \bar{k}_\perp^2}{a} + i\phi(\bar{k}_\perp)\right)\right]\delta(\bar{q} + \bar{D}) + c.c., \tag{8}$$

where c.c. indicates the complex conjugate of the last term and FT(I) is characterized by a strong peak at the origin and two sidebands centered at $q \approx \pm D$, as shown in Fig. 3e. If one of these sidebands is isolated, translated to the center of the $q$ plane and digitally inverse Fourier transformed, then the final complex image is

$$f(\bar{k}_\perp) = AB\ exp(i\frac{\pi^2 \bar{k}_\perp^2}{a} + i\phi(\bar{k}_\perp)). \tag{9}$$

The reconstructed phase is shown in Fig. 3f. The saturation of the colors represents the intensity, while the hue represents the phase. The phase is still affected by a parabolic term, which can be removed analytically. However, this term is rotationally symmetric and does not affect the characterization of the OAM. Corresponding simulations are shown for the ideal interferometry (Figs. 3g and 3h) and the reconstruction (Fig. 3i). Differences between the simulations and the experimental results arise from limitations of hologram fabrication.

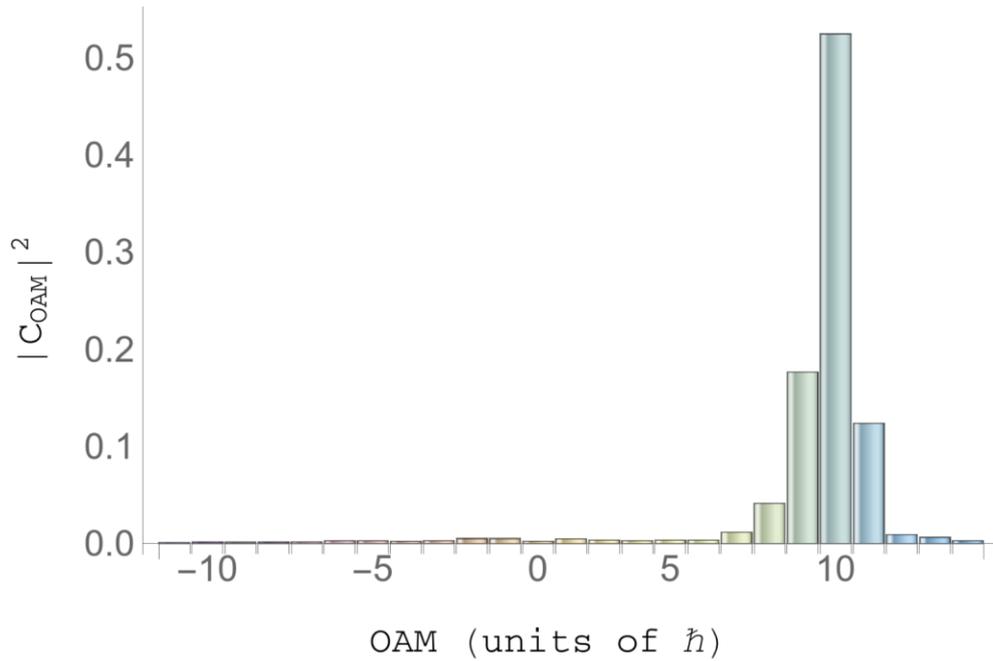

FIG. 4 . OAM decomposition of the retrieved phase shift. The spectrum has a narrow distribution centered on an OAM of 10ℏ

In Fig. 4, we show the OAM decomposition based on the reconstructed phase, as a benchmark of the reconstruction, obtained by transforming the phase in Fig. 3f to polar coordinates and Fourier transforming, as explained elsewhere [32]. The resulting narrow distribution is peaked at an OAM of 10ℏ as we aimed for. Whereas, in principle, the leading coefficient could reach a value of 80% [41], we obtain a slightly broadened distribution, almost certainly due to hologram imperfections or the limited coherence of the electron beam. Both of these effects can act as a low-pass filter and distort the reconstructed image. Nevertheless, the general reconstruction is reliable.

Although the application of the approach to an EVB is a case study, this method can be used in a more general context, especially for the diffraction of micron-sized objects ranging from magnetic materials to strained semiconductors and biological structures. In the present example, we used an approach based on a carrier frequency for both the object and the reference hologram, in order to achieve more precise phase control. However, holograms and objects without a carrier frequency can also be considered. It is also possible to push the maximum frequency towards nm$^{-1}$ scales by using a different setup of the reference "defocused beams", improved nanofabrication, and/or the use of holograms located in different apertures in the electron microscope column.

Outside electron microscopy, the concept of diffraction holography using a non-plane reference wave has also been proposed for X-rays [37,38]. However, the use of a pinhole to generate the reference wave then made the interference effect weak. In contrast, for electrons it is possible to equalize the intensities of the two beams to maximize the interference effects.

In summary, we have demonstrated the use of a defocused beam with a parabolic phase profile for object phase retrieval and applied it to an electron vortex beam. In electron microscopy, the technique permits the phases of diffraction images to be retrieved directly in Fraunhofer conditions, providing opportunities for large area interferometry and easy interpretation of the results without the influence of Fresnel fringes. Here, we could use the approach to confirm that a vortex beam has an intended average OAM value of $10\hbar$.

**Acknowledgments**

V.G. acknowledges the support of the Alexander von Humboldt Foundation. S.F. and F.V. acknowledge Unimore for support (FAR 2015). R.D.-B. is grateful to the to the Deutsche Forschungsgemeinschaft for a Deutsch-Israelische Projektkooperation (DIP) Grant. The research leading to these results has received funding from the European Research Council under the European Union's Seventh Framework Programme (FP7/2007-2013)/ ERC grant agreement number 320832. R.W.B. and E.K. acknowledge the support of Canada Excellence Research Charis (CRC) program.

[41] A characterization of OAM purity in our hologram as in ref 29 with the method in ref 31 led to a typical purity in OAM decomposition of 80%.